\documentclass[cits]{PoS}

\title{Hydrodynamics of Young Supernova Remnants and the Implications for their Gamma-ray emission }

\ShortTitle{Young Supernova Remnants and their Gamma-ray emission }

\author{\speaker{Vikram Dwarkadas}%
        University of Chicago\\
        E-mail: \email{vikram@oddjob.uchicago.edu}}

\abstract{Supernovae (SNe) are generally classified into Type I and
  Type II. Most SNe ($\sim$ 80\%), including all the subtypes of Type
  II, and Type Ib/c, arise from the core-collapse of massive
  stars. During their lifetime, mass-loss from these stars
  considerably modifies the medium around the stars. When the stars
  explode as SNe, the resulting shock wave will expand in this
  wind-modified medium. In contrast, Type Ia SNe will expand in a
  relatively uniform medium, but the dynamics are different from those
  of core-collapse SNe. For young supernova remnants, the properties
  of the ejecta as well as the surrounding medium are important in
  determining the subsequent evolution of the SN shock wave, and the
  dynamics and kinematics of the remnant. This will influence the
  acceleration of particles at the SN shocks, and consequently affect
  the gamma-ray emission from the remnant.\\

Herein we discuss the expected properties, especially the density
structure, of the medium around various types and sub-types of SNe, as
suggested by current stellar evolution models.  Using analytic and
semi-analytic models and numerical simulations, we investigate how
these affect the kinematics of the SN shock waves, assess the impact
this would have on the production of cosmic rays, and show how it
influences the time-evolution of the hadronic gamma-ray emission from
the remnant. In the case of SNRs evolving in a wind medium, the
emission should reach a maximum early on, and thereafter decrease with
time. For SNe in a constant density medium, the emission would be
expected to increase with time upto the advent of the Sedov stage.}

\FullConference{Cosmic Rays and the InterStellar Medium\\
		 24-27 June 2014\\
		 Montpellier, France}

\begin{document}

\section{Introduction}
Supernovae (SNe) are basically divided into two types - those that
arise from the thermonuclear detonation of low mass white dwarf stars
(the Type Ia SNe) and those that arise from the core-collapse of
massive stars, which include all the other types such as 1b, 1c, and
all Type II SNe. The explosion that forms either type of SN results in
the expansion of a fast shock wave into the surrounding medium. The
expansion and evolution of this shock depends on the characteristics
of both the ejected material from the SN (the SN `ejecta') and the
structure of the surrounding material. In particular the density
structure of both the ejecta and ambient medium are important. In this
paper we discuss the formation of the ambient medium around various
types of SNe into which the shock wave evolves, the resultant
evolution of the SN shocks within this medium, and the implications
for the gamma-ray emission from the various types of SNe.

\section{Young Supernova Remnants}
\label{sec:youngsnr}
We define young supernova remnants (SNRs) as those that are still in
the ejecta-dominated stage, and have not yet reached the Sedov-Taylor
phase. To reach the Sedov-Taylor phase, the SN shock front must sweep
up an amount of mass several tens of times larger than the ejected
mass \cite{dc98}, which may take hundreds of years, or even thousands
of years for a SNR evolving in a low density medium, such as SN
1006. The expansion of a SNR in the ejecta dominated phase leads to
the formation of a strong shock expanding into the ambient medium, and
a reverse shock that expands back into the ejecta in a Lagrangian
sense. The two are separated by a contact discontinuity that divides
the shocked ejecta from the shocked ambient medium.

\subsection{Type Ia Supernovae} 
Type Ia SNe are thought to arise from the deflagration or detonation
of a white dwarf star, presumably in a binary system. Since the white
dwarfs don't suffer from wind-driven mass-loss, we can assume, at
least in the first approximation, that they do not modify the medium
around them. Thus the medium can be taken to be the interstellar
medium, with a constant density profile. It is of course possible that
the companion star may modify the medium, at least close-in to the SN,
which is a possibility that needs to be examined more thoroughly.

By comparing to the ejecta profiles obtained from Type Ia explosions,
\cite{dc98} showed that the best approximation to the ejecta density
distribution is an exponential profile. This fits much better than a
power-law distribution that has often been used by many
authors. However the introduction of an exponential introduces an
additional parameter, and thus a self-similar solution is no longer
possible. As shown by \cite{dc98}, the radius of the forward shock
given by this profile is comparable to that given by a
power-law. However the radius of the reverse shock, and the density
profile within the shocked region, is quite different.

\begin{figure}
\includegraphics[width=1.\textwidth]{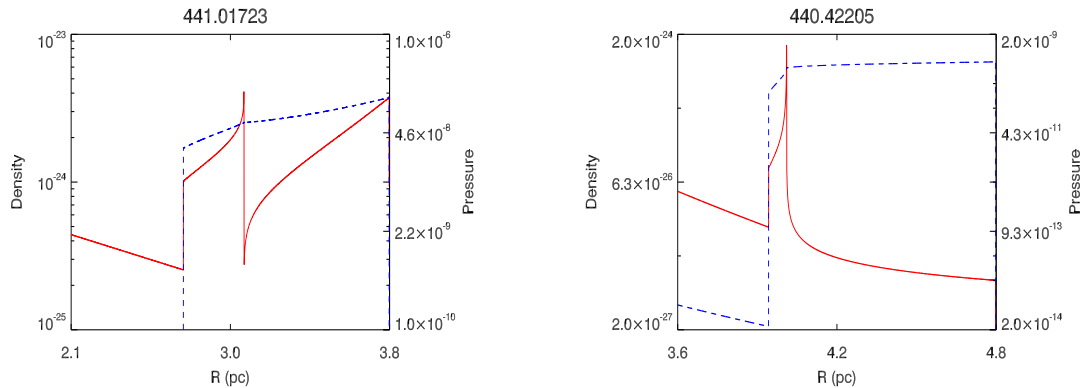}
\caption{(Left) The density and pressure profiles for a Type Ia SN,
  described by an exponential profile, expanding into a constant
  density medium.  (RHS) The profiles due to a Type II SN, described
  by a power-law profile, expanding into a wind-blown region. In both
  cases, the red line indicates the density (left axis scale), and
  blue the pressure (right axis scale). The radius is in parsecs, and
  time (top) is given in years. The grid is expanding, such that the
  outer edge of the grid lies just beyond the outer shock, which is
  therefore seen very close to the edge of the grid.}
\label{fig:snprofiles}
\end{figure}

The density and pressure profile of a Type Ia SNR expanding into a
constant density medium is shown on the left hand side of Figure
\ref{fig:snprofiles}.

\subsection{Core-Collapse SN}
The evolution of core-collapse SNe, and the nature of the medium
surrounding them, is more complicated. Core-collapse SNe arise from
massive stars (M $ \ge 8 {\rm M}_{\odot}$). These stars lose mass
continually in the form of stellar winds, and for some stars via mass
``eruptions''. The amount of mass lost can be a substantial fraction
of the stellar mass, and can modify the medium outside the star over a
radius of several tens of parsecs \cite{vvd05}, depending on the
surrounding density.

\begin{figure}
\includegraphics[width=1.05\textwidth]{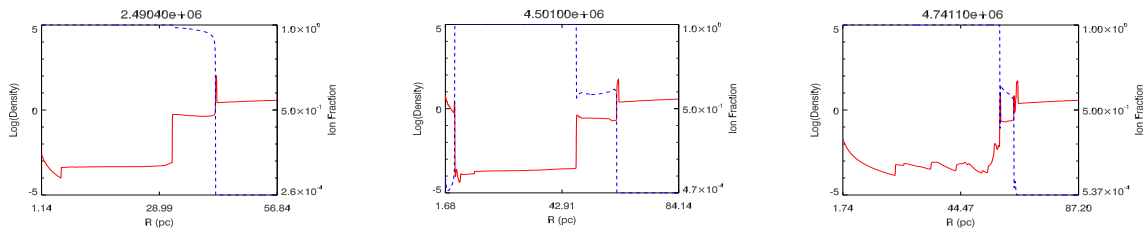}
\caption{Evolution of the medium around a 40 M$_{\odot}$ star. The
  star goes through three stages, the main sequence phase (left), the
  red supergiant phase (middle) and the Wolf-Rayet phase (right). The
  red line displays the logarithm of the number density (scale on
  left); the blue is ionization fraction (scale on right). Radius is
  in parsecs, and time (top) is in years. }
\label{fig:wrbub}
\end{figure}

The formation of the medium surrounding a massive star, due to the
stellar wind mass-loss combined with ionization from the star, has
been studied by many groups (e.g.~\cite{dr13, ta11}, and references
within). In Figure \ref{fig:wrbub} we show the evolution of the medium
around a 40 M$_{\odot}$ star over the stellar lifetime \cite{vvd07,
  dr13}. In the main sequence phase (left), which occupies most of the
star's lifetime, a wind bubble forms. Going outwards in radius, we see
a freely expanding wind, wind-termination shock, a low density region
of shocked wind, an ionized region, a contact discontinuity separating
the shocked surrounding medium from the bubble interior, and an outer
shock, which is generally radiative. When the star leaves the main
sequence to become a red supergiant (RSG), it blows a high mass-loss
rate and low velocity wind, which creates a new pressure distribution
and a high-density wind near the star. Eventually the star may lose
its hydrogen, and sometimes He, envelope, becoming a Wolf-Rayet (W-R)
star, and emitting a wind with a density that is a few times lower
than in the RSG stage, but a velocity that is two orders of magnitude
higher. This results in a W-R wind with a much higher momentum,
capable of pushing the RSG wind out and mixing its contents in the
bubble. At the same time, the density of the medium, which depends on
the ratio of wind mass-loss rate to wind velocity, is considerably
lower than in the RSG phase.

Although this is a simplified overview of one specific star, it
illustrates some general considerations:\\

\noindent
$\bullet$ Near a massive star, irrespective of which phase it is in,
the density profile resembles that of a freely expanding wind
medium. If the mass-loss rate and velocity are constant, this density
will decrease as r$^{-2}$. SNe from massive stars will initially
expand within this wind region.\\
$\bullet$ For stars that explode to form SNe when in the RSG phase
(Type IIP, IIL and possibly IIb SNe), the wind density around the star
will be quite high, due to the high mass-loss rate and low wind
velocity. Therefore these SNe will expand in a high density wind,
although the density may fall after several hundred years (Figure
\ref{fig:wrbub}, middle).\\ 
$\bullet$ For stars that explode to form SNe when in the W-R stage
(Type Ib/c), the wind density is much lower than for IIPs (Figure
\ref{fig:wrbub}, right). These stars are more compact than the RSGs as
they have lost their H and, in some cases, He envelopes. The fast
moving shock wave expands in a low density medium before colliding
with the wind termination shock.\\

\section{Evolution of Young SNRs} 
The evolution of a young SNR will depend on both its ejecta density
profile and the nature and structure of the surrounding medium.  As
described above, Type Ia SNRs can be thought of as evolving, at least
to zeroth order, in the constant density ISM. The crucial parameter is
thus the density of the medium. Core-collapse SNe will evolve, at
least initially, in the stellar wind, a medium of decreasing
density. Over time however the density profile will change, which will
directly affect the evolution. 

The evolution of the supsersonic ejecta into the surrounding medium
leads to a double-shocked structure, as described in \S
\ref{sec:youngsnr} (Figure \ref{fig:snprofiles}). The density
structure of the shocked material within the two shocks depends on
both the ejecta profile and the ambient medium, and will vary for
various types of remnants expanding into different media. This will
directly affect any emission process that is dependent on the
density. Thus accurate calculations of the emission require that the
hydrodynamics be properly calculated. If the cosmic-ray acceleration
is efficient, and the energy expended in accelerating particles is
high ($\ge 10$\%), then the back-reaction due to cosmic rays could
affect the hydrodynamics itself, changing the shock structure and the
relative positions of the forward shock, contact discontinuity and
reverse shock \cite{cspe11,fds12}.

\begin{figure}
\includegraphics[width=1.0\textwidth]{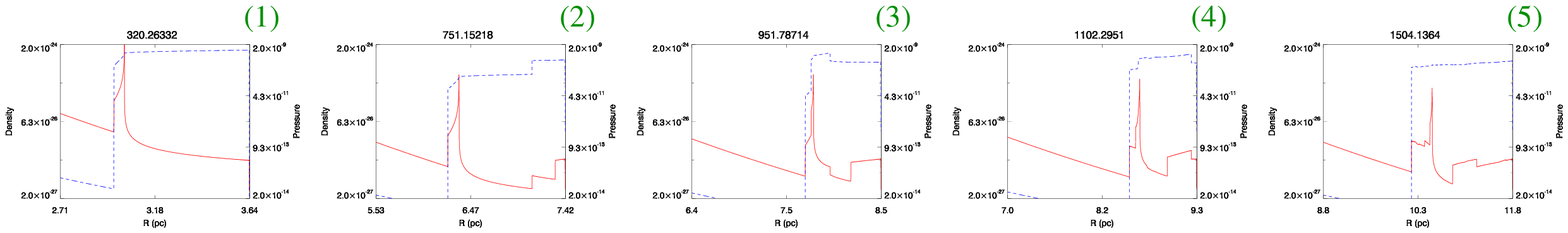}
\caption{Evolution of a SN shock wave within the wind bubble created
  by a W-R star. The simulation focuses on the region in between the
  two shocks. In successive frames we show the SNR evolving in the
  freely expanding wind (1); colliding with the wind termination
  shock, giving a reflected shock that goes back into the ejecta; (3)
  the reflected shock crosses the contact discontinuity, sending a
  rarefaction wave into the shocked ambient medium; (4) and (5) the
  impact of the transient shock waves reduces as the complicated
  structure continues to expand outwards.  The red line displays
  density (scale on left); the blue is pressure (scale on right). The
  radius is in parsecs, and time (top) is given in years. }
\label{fig:type1c}
\end{figure}

The SN shock waves interacting with the ambient medium can lead to a
complicated evolution, especially for core-collapse SNe. The evolution
has been studied by many authors (see \cite{vvd05, vvd07} and
references therein). In the case of core-collapse SNe expanding in a
low density wind-blown bubble (Figure \ref{fig:wrbub}, right), for
example, the SN shock is first interacting with a freely expanding
wind, then collides with the wind-termination shock, and then expands
into a more-or-less constant density medium (Figure \ref{fig:type1c}),
before eventually impacting the dense shell.

\noindent
$\bullet$ As shown in \cite{vvd05} (Figure\ref{fig:type1c}), the
impact of a strong shock on another shock wave or discontinuity leads
to a transmitted shock into the medium and a reflected shock wave back
into the ejecta. Thus, besides the SN shocks, there can be other shock
waves in an inhomogeneous medium, which may at least temporarily
accelerate particles to high energies. The observed spectrum is a
composite of all of these various interactions, and should be
appropriately computed. \\ $\bullet$ The density structure of the
shocked material changes considerably as the SN moves from a wind
medium to a constant density medium, which would affect the emission
from the remnant. \\ $\bullet$ The shock velocity on impact with the
wind termination shock is reduced, and then slowly increases again,
which can affect the acceleration of particles.\\

It is important to realize (Figure \ref{fig:type1c}) that even when a
SNR is expanding in a wind-blown bubble, it need not be expanding in a
freely expanding wind with a density profile that decreases as
r$^{-2}$, but could transition to a shocked wind with approximately
constant density (with some fluctuations). The shock velocity and
density structure will be modified. Furthermore, even when in the
constant density region, the density profile may not resemble that it
would have if it had originally started expanding in the constant
density region \cite{vvd05}. This has implications for instance in
modelling the $\gamma$-ray emission from SNR RXJ1713, assumed to arise
within a low density wind bubble \cite{espb12}.

\section{Gamma-Ray emission} 
Supernova shock waves can accelerate particles to high energies,
presumably by the process of Diffusive Shock Acceleration
\cite{drury83}. The acceleration process is highly dependent on the
magnetic field and magnetic turbulence ahead of the shock. This leads
to a question, as yet unanswered, of whether the reverse shock is
capable of accelerating particles, since the magnetic field in the SN
ejecta is unknown. There is however no question that the forward shock
can accelerate particles. The maximum energy of the particles depends
on the magnitude of the magnetic field, and its orientation. Magnetic
fields estimated from radio observations appear to far exceed the
interstellar field, and it has been suggested that the back reaction
of the cosmic rays themselves may amplify the field
\cite{bl01}. Although the actual mechanism remains unidentified,
several instabilities have been identified that may serve to amplify
the field. In a constant density medium, such as for Type Ia SNe, the
maximum energy of the accelerated particles will initially increase
with time, although it will begin to decrease in most cases before the
remnant has reached the Sedov stage \cite{dtp12, sb13}. In a wind
medium, maximum energies on the order of a PeV could be reached in a
few days \cite{mrdt14}, after which the energy of the particles would
decrease with time.

The acceleration of protons to high energies, and their collision with
stationary ambient protons, leads to hadronic emission via pion
decay. For a SN expanding in a RSG wind (Figure \ref{fig:wrbub},
middle), the SN shock is first expanding in a high density RSG wind,
but later will continue to expand in a low density medium created by
the main sequence \cite{tdp13}. At some point the forward shock will
begin to expand within the low density medium while the reverse shock
will still be in the high density wind. If the reverse shock is
accelerating particles to very high energies, then hadronic emission
would predominate at the reverse shock, whereas accelerated electrons
may be producing leptonic emission at the forward shock. The total
$\gamma$-ray emission will be a combination of leptonic and hadronic
processes, and not attributable to a single process. A combination of
processes is possible in many other scenarios.

Using a simple analytic model for the hadronic emission \cite{dav94},
and the Chevalier self-similar solution \cite{rac82} for the expansion
of a young SNR into the surrounding medium, \cite{vvd13} calculated
the evolution of the gamma-ray flux with time for a SNR expanding in a
medium with a power-law profile.  In the case of a wind medium with
constant mass-loss parameters, the flux is given by:

\begin{eqnarray}
F_{\gamma} (> E_o,t) & = & \frac{3 q_{\gamma} B^2 \xi {(\kappa C_1)} m^3 }{2(5m-ms-2)\beta \mu m_p d^2} t^{m-2} \\
& = & \frac{3 q_{\gamma} \xi {(\kappa C_1)} m^3 }{32 {\pi}^2 (3m-2)\beta \mu m_p d^2}\left[\frac{\dot{M}}{v_w}\right]^2 t^{m-2}
\label{eq:pionwind}
\end{eqnarray}

In the above equation, the contact discontinuity expands as R$_{CD} =
C_1 t^m$, and the radius of the forward shock as $R_{sh} = \kappa
R_{CD} = \kappa C_1 t^m$. The velocity $v_{sh} = d\,R_{sh}/dt = m
\kappa C_1 t^{m-1}$ is therefore always decreasing with
time. $q_{\gamma}$ is the $\gamma$-ray emissivity normalized to the
cosmic-ray energy density (tabulated in \cite{dav94}); $\xi$ denotes
the fraction of the shock energy that is converted to cosmic rays; $d$ is the distance to the source; $\mu$ is the
mean molecular weight and $m_p$ is the proton mass; $\dot{M}$ is the
wind mass-loss rate and $v_w$ is the wind velocity; and $\beta \sim
0.3-0.5 $ accounts for the fact that the volume of the shocked region
from which the emission arises is smaller than the volume of the
entire SNR.

The time dependence of the $\gamma$-ray flux is t$^{m-2}$. Since for a
wind-blown medium, $2/3 < m \le 1$ in the ejecta-dominated stage, this
indicates that the flux is decreasing with time. Thus one would expect
the flux to reach a maximum energy in a short period of time, and
decrease from there on. The best time to observe SN in winds
(essentially all core-collapse SNe) would then be soon after
explosion, when the flux is near maximum (see also \cite{mrdt14}).

We can apply this to Cas A \cite{vvd13}, assuming that it is expanding
in a dense RSG wind, using the parameters in \cite{co03}. The {\it
  Fermi} best fit suggests that at most 2\% of the total energy has
gone into cosmic rays. If we assume $\xi=0.02$ then we get a flux
${F_{\gamma}}_{CAS A} (> 100 MeV) = 1.1 \times 10^{-8} $ that is
comparable to the {\it Fermi} result. This suggests that a hadronic
description may fit Cas A.

For a SNR expanding in a constant density medium, \cite{vvd13} found
the flux to be given as:

\begin{equation} 
F_{\gamma} (> E_o,t) = \frac{3 q_{\gamma} \xi {(\kappa C_1)^5} m^3
}{6 (5m-2)\beta \mu m_p d^2} \rho_{am}^2 t^{5 m-2}
\label{eq:pionconst}
\end{equation}

For a SNR in a constant density medium, the parameter $m=2/5$ in the
Sedov phase, and $m > 2/5$ in the ejecta-dominated phase.  Thus in a
constant density medium the hadronic emission is increasing with time
in the ejecta-dominated stage. Although this has been derived using a
power-law profile, we expect a similar dependence for an exponential
ejecta profile. This then suggests that core-collapse and Type Ia SNe
show a different time-dependence in their hadronic emission, with the
former decreasing with time and the latter increasing with time. 

\begin{figure}
\includegraphics[width=1.0\textwidth]{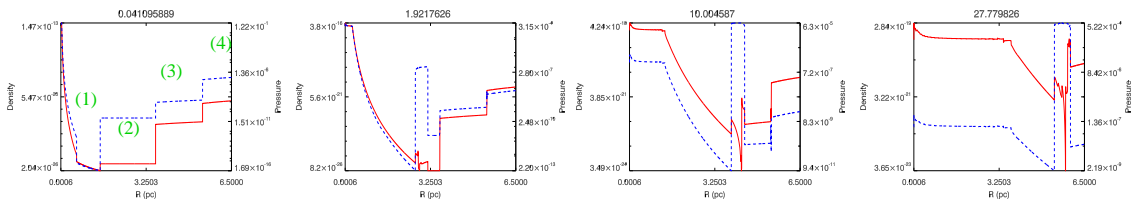}
\caption{Evolution of the SN shock wave in the medium surrounding SN
  1987A \cite{deweyetal12}. The medium consists of 4 regions (1) The
  freely expanding Blue Supergiant Wind (2) The Shocked Wind (3) The
  HII region and (4) The dense shell (equatorial ring), including the
  ``fingers''responsible for the bright hotspots. The time evolution
  of the shock in regions (2), (3) and (4) is shown. }
\label{fig:87a}
\end{figure}

If a core-collapse SN were to expand in a constant density medium, its
flux would increase with time. This is the expected case with SN
1987A, which is expanding into the high density HII region and dense
shell created by the interaction of winds from the progenitor blue
supergiant star \cite{cd95}. The nature of the medium, and the
evolution of the SN shock wave within this medium, is shown in Figure
\ref{fig:87a}, using the parameters in \cite{deweyetal12}. Using
equation \ref{eq:pionconst} with the appropriate parameters
\cite{deweyetal12}, we find \cite{vvd13} (a) the flux is increasing
with time, as expected, and (b) that in another 5-10 years, as the
shock wave sweeps up more material, the flux will be large enough to
be detectable potentially by the HESS array, and with high probability
by the upcoming Cerenkov Telescope Array.

It is also useful to note the quadratic dependence on the density, as
opposed to the linear dependence in \cite{dav94}. While one power of
density arises from the target mass, as in \cite{dav94}, the second
power comes from the fact that the energy available to be extracted at
the shock front is not a constant, but is in fact increasing with
time. In the above $\xi$ is assumed to be constant; however it is
quite possible that it could be a function of time.

{\bf Acknowledgements} VVD's research on very high energy emission
from young SNRs is partially supported by NASA Fermi grant NNX12A057G,
and has benefited considerably from interaction with A.~Marcowith,
M.~Pohl, M.~Renaud, V.~Tatischeff, and I.~Telezhinsky.  We are
grateful to the FACCTS program for supporting travel to France, and
fostering collaboration with University of Montpellier II. We thank
A.~Marcowith for hosting a very interesting and stimulating
conference.


\begin{thebibliography}{99}
  \bibitem{bl01} A. R. Bell, and S. G. Lucek, ~\emph{Cosmic ray
    acceleration to very high energy through the non-linear
    amplification by cosmic rays of the seed magnetic field},
    \emph{MNRAS} {\bf 321}, (2001), 433
    \bibitem{cspe11} D.~Castro, P.~Slane, D.~Patnaude and D.~Ellison,
      \emph{The Impact of Efficient Particle Acceleration on the
        Evolution of Supernova Remnants in the Sedov-Taylor Phase},
      \emph{ApJ} {\bf 734}, (2011), 85
      \bibitem{co03} R.~A.~Chevalier, and J.~Oishi, \emph{Cassiopeia A
        and Its Clumpy Presupernova Wind}, \emph{ApJ} {\bf 593},
        (2003), 23
     \bibitem{cd95} R. A. Chevalier, and V.~V.~Dwarkadas, \emph{The
       Presupernova H II Region around SN 1987A}, \emph{ApJL} {\bf
       452}, (1995), L45
   \bibitem{rac82} R. A. Chevalier,~\emph{Self-similar solutions for
     the interaction of stellar ejecta with an external medium},
     \emph{ApJ} {\bf 258}, (1982), 790
    \bibitem{dav94} L.~O'C.~Drury, F.~A.~Aharonian, and
      H.~J.~Voelk,~\emph{The gamma-ray visibility of supernova
        remnants. A test of cosmic ray origin}, \emph{A\&A}, {\bf
        287}, (1994), 959, [{\tt astro-ph/9305037}].
    \bibitem{deweyetal12} D.~Dewey, V.~V.~Dwarkadas, F.~Haberl,
      R.~Sturm, and C.~R.~Canizares,~\emph{Evolution and Hydrodynamics
        of the Very Broad X-Ray Line Emission in SN 1987A}, \emph{ApJ}
      {\bf 752}, (2012), 103, [{\tt arXiv:1111.5314}].
  \bibitem{drury83} L.~Oc.~Drury, ~\emph{An introduction to the theory
    of diffusive shock acceleration of energetic particles in tenuous
    plasmas}, \emph{RPP}, {\bf 46}, (1983), 973
    \bibitem{vvd13} V.~V.~Dwarkadas, ~\emph{Exploring the $\gamma$-ray
      emissivity of young supernova remnants - I. Hadronic emission},
      \emph{MNRAS}, {\bf 434}, (2013), 3368, [{\tt arXiv:1307.4414}].
  \bibitem{dr13} V.~V.~Dwarkadas and D.~Rosenberg, ~\emph{Simulated
    X-ray spectra from ionized wind-blown nebulae around massive
    stars}, \emph{HEDP}, {\bf 9}, (2013), 226, [{\tt arXiv:1206.1348}].
    \bibitem{dtp12}V.~V.~Dwarkadas, I.~Telezhinsky, and M.~Pohl,
      \emph{On the maximum energy and escape of accelerated particles
        in young supernova remnants}, in proceedings of \emph{HIGH
        ENERGY GAMMA-RAY ASTRONOMY: 5th International Meeting},
      \emph{AIPC}
   \bibitem{vvd07} V.~V.~Dwarkadas, ~\emph{The Evolution of Supernovae
     in Circumstellar Wind Bubbles. II. Case of a Wolf-Rayet Star},
     \emph{ApJ}, {\bf 667}, (2007), 226, [{\tt arXiv:0706.1049}].
  \bibitem{vvd05} V.~V.~Dwarkadas, ~\emph{The Evolution of Supernovae
    in Circumstellar Wind-Blown Bubbles. I. Introduction and
    One-Dimensional Calculations}, \emph{ApJ}, {\bf 630}, (2005), 829,
    [{\tt astro-ph/0410464}].
  \bibitem{dc98} V.~V.~Dwarkadas and R.~Chevalier, ~\emph{Interaction
    of Type IA Supernovae with Their Surroundings}, \emph{ApJ}, {\bf
    497}, (1998), 807
    \bibitem{espb12} D.~Ellison, P.~Slane, D.~Patnaude, and A.~Bykov,
      \emph{Core-collapse Model of Broadband Emission from SNR RX
        J1713.7-3946 with Thermal X-Rays and Gamma Rays from Escaping
        Cosmic Rays}, \emph{ApJ}, {\bf 744}, (2012), 39, [{\tt
          arXiv:1109.0874}].
    \bibitem{fds12} G.~Ferrand, A.~Decourchelle, and S.~Safi-Harb,
      \emph{3D Simulations of the Thermal X-Ray Emission from Young
        Supernova Remnants Including Efficient Particle Acceleration},
      \emph{ApJ}, {\bf 760}, (2012), 34, [{\tt arXiv:1210.0085}].
  \bibitem{mrdt14} A.~Marcowith, M.~Renaud, V.~V.~Dwarkadas,
    V.~Tatischeff, ~\emph{Cosmic-ray acceleration and gamma-ray
      signals from radio supernovae}, in proceedings of \emph{Cosmic
      Ray Origin beyond the standard models}, [{\tt arXiv:1409.3670}].
    \bibitem{sb13} K.~M.~Schure, and A.~R.~Bell, \emph{Cosmic ray
      acceleration in young supernova remnants}, \emph{MNRAS}, {\bf
      435}, {2013}, {1174}, [{\tt arXiv:1307.6575}].
   \bibitem{tdp13}I.~Telezhinsky, V.~V.~Dwarkadas, and M.~Pohl,
     \emph{Acceleration of cosmic rays by young core-collapse
       supernova remnants}, \emph{A\&A}, {\bf 552}, (2013), {102},
          [{\tt arXiv:1211.3627}].
 
    \bibitem{ta11} J.~A.~Toala, and S.~J.~Arthur,
      \emph{Radiation-hydrodynamic Models of the Evolving
        Circumstellar Medium around Massive Stars}, \emph{ApJ}, {\bf
      737}, (2011), {100}, [{\tt arXiv:1106.4493}].

\end{thebibliography}
\end{document}